\documentclass[12pt]{article}
\usepackage{amsmath}
\usepackage{amsbsy}
\usepackage{amssymb}
\usepackage{color}

\newcommand{\beq}{\begin{equation}}
\newcommand{\eeq}{\end{equation}}
\newcommand{\bdm}{\begin{displaymath}}
\newcommand{\edm}{\end{displaymath}}
\newcommand{\beqr}{\begin{eqnarray}}
\newcommand{\eeqr}{\end{eqnarray}}
\newcommand{\beqrn}{\begin{eqnarray*}}
\newcommand{\eeqrn}{\end{eqnarray*}}

\def\l{\lambda}
\def\bchi{\boldsymbol{\chi}}
\def\M{\boldsymbol{M}}
\def\k{\kappa}
\def\l{\lambda}

\begin{document}

\title{A note on Franel numbers and $SU(3)$}

\author{Jos\'e Fern\'andez N\'u\~{n}ez$^{\dagger}$, 
Wifredo Garc\'{\i}a Fuertes$^{\ddagger}$
\\
\small Departamento de F\'\i sica, 
Facultad de Ciencias, Universidad de Oviedo, 33007-Oviedo, Spain\\
\small {\it $^\dagger$nonius@uniovi.es; 
$^\ddagger$wifredo@uniovi.es}\\
\and
 Askold M. Perelomov\\
\small Institute for Theoretical and Experimental Physics, Moscow, Russia.\\
\small {\it  aperelomo@gmail.com}}

\date{ }

\maketitle

\begin{abstract}
\noindent
We study weight multiplicities in tensor powers of the adjoint representation of the Lie algebra of $SU(3)$ and, by means of these multiplicities and the theory of the quantum integrable Calogero-Sutherland model, we develop a procedure to reduce some sums of triple products of binomial coefficients to linear combinations of Franel numbers. A new way to prove the Franel recurrence relation based on these results is presented.
\end{abstract}
\bigskip

%{\bf PACS:} 02.30.Ik,  02.20.Qs,  03.65.Fd.


\medskip

\vfill\eject 
\noindent{\large{\bf 1.}} The sums
\bdm
S_n^{(r)}=\sum_{k=0}^n \binom{n}{k}^r
\edm
for $r=1,2$ are easy to compute and they satisfy simple recurrence relations,
\bdm
S_{n+1}^{(1)}-2 S_{n}^{(1)}=0,\hspace{2cm}(n+1)S_{n+1}^{(2)}-2(2n+1) S_{n}^{(2)}=0,
\edm
but higher $r$ cases are not so straightforward. In 1894, J. Franel discovered a three-term recurrence relation for the case $r=3$, 
\beq
(n+1)^2 S_{n+1}^{(3)}-(7 n^2+7 n+2)S_n^{(3)}-8 n^2 S_{n-1}^{(3)}=0,\label{franrec}
\eeq
and a year later he found another one for the case $r=4$ \cite{fr94}, proposing also the conjecture that there are recurrence relations with polynomial coefficients and $[\frac{1}{2}(r+3)]$ terms  for all $r$. This was corroborated for the cases $r=5$ and $6$ in 1987 by Perlstadt \cite{perl87} by means of a ``creative telescoping" technique using MACSYMA, and in 1989 Cusick \cite{cu89} presented a method, based on the intertwining of two identities involving the sum of some powers of combinatorial numbers, which can be used to find the recurrence relations for higher $r$. Our purpose in this note is to come back to the original $r=3$ case, in this occasion dealing with it under the optic of Lie algebra representation theory. Namely, we will compute the multiplicities of the weights entering in the $n$-th  tensor power of the adjoint representation of $SU(3)$ and figure out how to express them in terms of the Franel numbers $F_n\equiv S_n^{(3)}$. This perspective brings about some nontrivial relations among these numbers and other sums of cubic products of binomial coefficients, and it gives also the opportunity of presenting a new proof of (\ref{franrec}) based on Lie algebra methods. Although we will confine here the treatment to $r=3$, we think that similar considerations applied to $SU(r)$ for $r>3$ will yield further nontrivial relations of the same type and additional proofs of the Franel recurrence relations for higher $r$.\\

\noindent{\large{\bf 2.}} We begin by very briefly recalling a few facts about the quantum Calogero-Sutherland model associated to the $SU(3)$ Lie algebra that will turn useful below. The Calogero-Sutherland models related to simple Lie algebras, see \cite{op83} for a detailed review, arise in mathematical physics as a class of completely integrable systems which describe the classical or quantum-mechanical interaction of several particles, where the forces are derived from a potential built from the root system of the   algebra. Specifically, for the case of $SU(3)$ there are three identical particles, with coordinates $q_1,q_2$ and $q_3$, and the potential, which depends on a single coupling constant $\kappa$, is of the form
\bdm
U(q_1,q_2,q_3)=\kappa(\kappa-1) \left(\sin^{-2}(q_1-q_2)+\sin^{-2}(q_1-q_3)+\sin^{-2}(q_2-q_3)\right),
\edm
while the center of mass is fixed at the origin of coordinates,  i.e. $\sum_iq_i=0$. In the quantum case, after solving the Schr\"{o}dinger equation, one finds \cite{pe98,prz98} that the quantum states are in one-to-one correspondence with the dominant weights of the $SU(3)$-Lie algebra, i.e. to the weights $\mu$ of the form $m_1\lambda_1+m_2\lambda_2$, with $m_1,m_2 \geq 0$ and $\lambda_1, \lambda_2$ the fundamental weights; see for instance \cite{fh91} for a general reference on Lie algebra representation theory.  In fact, the wave function corresponding to the dominant weight $m_1\lambda_1+m_2\lambda_2$ is proportional to a generalized Gegenbauer polynomial $P_{m_1,m_2}^\kappa(z_1,z_2)$ in two variables, which is an eigenfunction of an effective Hamiltonian
 \beq
\Delta^\kappa=(z_1^2-3 z_2)\partial_{z_1}^2+(z_2^2-3 z_1)\partial_{z_2}^2+(z_1 z_2-9)\partial_{z_1}\partial_{z_2}+(3\kappa+1)(z_1\partial_{z_1}+z_2\partial_{z_2})\label{lap}
 \eeq
derived from the original one $H=\frac12p^2+U$. The relation with the $q$-coordinates is as follows: $z_1=x_1+x_2+x_3$ and $z_2= x_1 x_2+x_1 x_3+x_2x_3$, where $x_k=e^{2 i q_k}$ and the localization of the center of mass implies the constraint  $x_1 x_2 x_3=1$. Thus, $z_1$ and $z_2$ are invariant under the Weyl  group  of $SU(3)$, and they can indeed be identified as the characters of the fundamental and anti-fundamental representations of the Lie algebra. The eigenvalue equation is\footnote{As a matter of notation, we will use also the short form $X_{m_1,m_2}$ for a generic function  $X_\mu$ defined by the weight $\mu=m_1\l_1+m_2\l_2$. }
\beq
\Delta^\kappa P_{m_1,m_2}^\kappa=\varepsilon_{m_1,m_2}(\kappa) P_{m_1,m_2}^\kappa\,,\hspace{1cm}\varepsilon_{m_1,m_2}(\kappa)=m_1^2+m_2^2+m_1m_2+3\kappa (m_1+m_2)\,,\label{sch}
\eeq
and the polynomials $P_{m_1,m_2}^\kappa$ can be obtained efficiently by solving it by a recursive procedure. These polynomials display many remarkable properties. In particular, $P_{m_1,m_2}^1$ is the character of the irreducible representation of $SU(3)$ with highest weight $m_1\l_1+m_2\l_2$, while $P_{m_1,m_2}^0$ coincides with the monomial symmetric function associated to the Weyl orbit of the same weight. Other values of $\k$ give interesting functions too. For instance, for $\k=\frac{1}{2},2,4$ they are zonal polynomials of $A_2$ type. Due to these properties, quantum Calogero-Sutherland models are a convenient tool to obtain a variety of results on the representation theory of simple Lie algebras, see for instance \cite{fgp18} or \cite{fgp14} and references therein. They have also many other applications in areas of both mathematics and physics \cite{vdi00,pol06}.

We shall mention for later use another property of the polynomials  $P_{m_1,m_2}^\kappa$: their derivatives with respect to the $z_1$-variable have a simple expression if the coupling constant is shifted by one, namely
\beq
\partial_{z_1} P_{m_1,m_2}^\kappa=m_1 P_{m_1-1,m_2}^{\kappa+1}+ A_{m_1,m_2}(\kappa) P_{m_1-2,m_2-1}^{\kappa+1}+B_{m_1,m_2}(\kappa) P_{m_1,m_2-2}^{\kappa+1}\label{deri}\,,
\eeq
with
\beqrn
A_{m_1,m_2}(\kappa)&=&\frac{m_1(m_1-1)m_2(m_1+m_2+\kappa-1)(m_1+m_2+\kappa)}{(m_1+\kappa-1)(m_1+\kappa)(m_1+m_2+2\kappa-1)(m_1+m_2+2\kappa)}\\
B_{m_1,m_2}(\kappa)&=&-\frac{m_2(m_2-1)(m_1+m_2+\kappa)}{(m_2+\kappa-1)(m_2+\kappa)}\,;
\eeqrn
there is a similar formula for the other derivative, see \cite{gp02} for the proofs. \\ 

\noindent{\large{\bf 3.}} Let us now consider tensor powers of the adjoint representation  $R_{\lambda_1+\lambda_2}$ of the $SU(3)$ Lie algebra. Specifically, we are interested in computing the multiplicity $a_\mu(n)=a_{p,q}(n)$ of each dominant weight $\mu=p\lambda_1+q\lambda_2$ entering in $R_{\lambda_1+\lambda_2}^{\otimes n}$ and, to this end, we will rely upon the use of the monomial symmetric functions defined by such weights,  which are given by 
\beq
\M_{p,q}=\sum_{{\cal P}\in{\cal S}_3} x_1^{{\cal P}(u_1)} x_2^{{\cal P}(u_2)} x_3^{{\cal P}(u_3)},\label{monomio}
\eeq
where ${\cal S}_3$ is the symmetric group of order three and the permutations act on  the elements of the  3-tuples $\bar u=(u_1,u_2,u_3)=(p+q,q,0)$, with the proviso that the constraint $x_1 x_2 x_3=1$ has to be respected. As it is well known, these permutations reproduce the action of the Weyl group of $SU(3)$ on the weight $p\l_1+q\l_2$, and thus the sum extends to all the weights of its orbit. The monomial symmetric functions form a basis of the space of functions invariant under the Weyl group of $SU(3)$, and the multiplicity we are looking for is the coefficient of $\M_\mu$ in the expansion in that basis of the $n$-th power of the character $\bchi_{1,1}$  of the adjoint representation of the algebra, i.e.,
\bdm
\bchi_{1,1}^n=\sum_{\nu\in \Lambda^+} a_\nu(n) \M_\nu,
\edm
where $\Lambda^+$ is the cone of positive weights and the explicit form of $\bchi_{1,1}$ is
\beq
\bchi_{1,1}=z_1 z_2-1=(x_1+x_2)(x_1+x_3)(x_2+x_3).\label{charadj}
\eeq 
Given that
\beq
\bchi_{1,1}^n=\sum_{k_1=0}^n \sum_{k_2=0}^n \sum_{k_3=0}^n \binom{n}{k_1} \binom{n}{k_2} \binom{n}{k_3} x_1^{n+k_1-k_3} x_2^{n+k_2-k_1} x_3^{n+k_3-k_2}\label{newton}
\eeq
and that we can write $\M_{p,q}=x_1^{p+q} x_2^q (x_1 x_2 x_3)^r+\ldots$ for any non-negative integer $r$, the Weyl symmetry of $\bchi_{1,1}^n$ allows us to identify $a_{p,q}(n)$ with the coefficient of $x_1^{p+q+r} x_2^{q+r} x_3^r$ in the expansion (\ref{newton}). Matching the exponents in this product with those in (\ref{newton}), we find a system of three equations for the $k_j$. The sum of these equations gives the compatibility condition $p+2q+3r=3n$,  which fixes $r$ and requires, in particular, that $p+2q$ is a multiple of 3. Once $r$ is fixed, we can solve for $k_1,k_2$ and $k_3$, obtaining $k_1=m$, $k_2=m+q-l$ and $k_3=m+q-2l$, where $l=\frac{p+2q}{3}$ and $m$ is any non-negative integer such the three $k_j$ remain between 0 and $n$. Finally, adopting the usual convention that binomial coefficients whose lower entry is negative or greater than the upper-one are zero, the final result can be written as follows:
\bdm
a_{p,q}(n)=0\hspace{0.5cm}{\rm if}\hspace{0.5cm} l=\frac{p+2q}{3} \notin {\bf Z}^+,
\edm
and if $l=\frac{p+2q}{3} \in {\bf Z}^+$ and $p\geq q$
\beq
a_{p,q}(n)=\sum_{j=0}^n \binom{n}{j} \binom{n}{j+q-l}\binom{n}{j+q-2l},\label{multi}
\eeq
while we can use $a_{q,p}(n)=a_{p,q}(n)$ in the other case.

It is also useful to work out the multiplicity $b_\mu(n)$ of the irreducible module $R_\mu$ in the tensor power $R_{\lambda_1+\lambda_2}^{\otimes n}$. The irreducible characters provide a different basis for Weyl-symmetric functions, and this new multiplicity corresponds to the coefficient of $\bchi_\mu$ of the expansion of $\bchi_{1,1}^n$ in that basis:
\beq
\bchi_{1,1}^n=\sum_{\nu\in \Lambda^+} b_\nu(n) \bchi_\nu\,.\label{expchar}
\eeq
The matrix elements for the change of basis between the $\M_\nu$ and the $\bchi_\rho$ are the so-called Kostka numbers, and in the present case they can be found as follows. The Weyl character formula for $SU(3)$ is
\beq
\bchi_{p,q}=\frac{\psi_{p,q}}{\psi_{0,0}}\,,\hspace{1cm} \psi_{p,q}=\left|\begin{array}{ccc}x_1^{p+q+2}&x_1^{q+1}&1\\x_2^{p+q+2}&x_2^{q+1}&1\\x_3^{p+q+2}&x_3^{q+1}&1\end{array}\right|,\label{weyl}
\eeq
and therefore, we can write
\beq
\psi_{0,0}^2 \bchi_{1,1}^n= \sum_{\mu\in \Lambda^+} b_\mu(n) \psi_{0,0} \psi_\mu\,.\label{char1}
\eeq
Both $\psi_{0,0}^2$ and $\psi_{0,0} \psi_\mu$ are Weyl-symmetric functions and can hence be expanded in the basis of the symmetric monomial functions. We will use the notation
\beq
\psi_{0,0}^2=\sum_{\nu\in \Lambda^+} (\psi_{0,0}^2)_\nu \M_\nu\,,\hspace{1cm} \psi_{0,0} \psi_\mu=\sum_{\nu\in \Lambda^+} (\psi_{0,0} \psi_\mu)_\nu \M_\nu\,.\label{quad} 
\eeq
Thus, if we have a rule
\bdm
\M_{\nu_1}\cdot \M_{\nu_2}=\sum_{\alpha\in \Lambda^+} C_{\nu_1,\nu_2}^\alpha \M_\alpha
\edm
to decompose the product of monomial symmetric functions, we can extract from (\ref{char1}) an infinite set of linear equations, one for each weight $\alpha$:
\beq
\sum_{\nu_1,\nu_2\in \Lambda^+} C_{\nu_1,\nu_2}^\alpha (\psi_{0,0}^2)_{\nu_1} a_{\nu_2}=\sum_{\mu\in \Lambda^+}(\psi_{0,0} \psi_\mu)_\alpha b_\mu(n)\,.\label{equ1}
\eeq
These equations can be solved iteratively to obtain the  multiplicities $b_\mu(n)$  in terms of the  $a_\nu(n)$ by virtue of the existence of a partial ordering among the weights entering in $\bchi_{1,1}^n$: $\mu_1$ is higher than $\mu_2$ if $\mu_1-\mu_2$ is a positive root. A list of the first weights in order of increasing height is as follows:
\beqrn
(p,q)&=&(0,0),(1,1),(3,0),(0,3),(2,2),(4,1),(1,4),(6,0),(0,6),(3,3),(5,2),\\
&&(2,5),(7,1),(1,7),(4,4),(6,3),(3,6),(5,5),(9,0),(0,9),(8,2),(2,8),\ldots 
\eeqrn

As for the rule providing the $C_{\nu_1,\nu_2}^\alpha $ coefficients, it can be formulated graphically in terms of Young diagrams \cite{fh91} or, equivalently, in algebraic form by means of 3-tuples. In this second version, we multiply the monomial functions as given in (\ref{monomio}), but including suitable numerical factors to take into account that they can have one, three or six terms. After these factors are properly adjusted, the rule reads as follows:
\beq
\M_{p,q}\cdot \M_{r,s}=\frac{1}{G_{p,q} G_{r,s}}\sum_{\bar{u}\in {\cal C}} G_{\mu(\bar{u})}\M_{\mu(\bar u)}\,,\label{product}
\eeq 
where $\bar{u}$ are 3-tuples of non-negative integers; the factors are: $G_{\mu(\bar u)}$ if the three entries in $\bar u$ are different, $G_{\mu(\bar u)}=2$  if only two entries coincide and $G_{\mu(\bar u)}=6$ if the three elements are equal; the weight corresponding to a 3-tuple is $\mu(\bar u)=(u_+-u_0)\l_1+(u_0-u_-)\l_2$, where  $u_+\geq u_0\geq u_-$ is the ordering of three elements of the tuple, and the sum extends to the set 
\beqrn
{\cal C}&=&\left\{(p+q+r+s,q+s,0),(p+q+s,q+r+s,0),(p+q+r+s,q,s),\right.\\
&&(p+q+s,q,r+s),(p+q,q+r+s,s),(p+q,q+s,r+s)\left.\right\}.
\eeqrn
To work with (\ref{equ1}) we also need the explicit form of the expansions (\ref{quad}), which can be obtained by hand by using (\ref{weyl}) and identifying the symmetric monomial functions appearing in these quadratic expressions. After some computations,  we find
\beqrn
\psi_{0,0}^2&=&-6 \M_{0,0}+2 \M_{1,1}+\M_{2,2}-2\M_{3,0}-2 \M_{0,3}\,,\\
\psi_{0,0} \psi_{p,q}&=&\widetilde{\M}_{p+2,q+2}-\widetilde{\M}_{p+3,q}-\widetilde{\M}_{p,q+3}-\widetilde{\M}_{p,q}+\varepsilon_{p-1}\widetilde{\M}_{p-1,q+2}+\varepsilon_{q-1}\widetilde{\M}_{p+2,q-1}\\&+&\delta_{p,0}\widetilde{\M}_{1,q+1}+\delta_{0,q}\widetilde{\M}_{p+1,1}\,,
\eeqrn
where $\widetilde{\M}_{c,d}=G_{c,d} \M_{c,d}$ and $\varepsilon_r$ is one if $r\geq 0$ and zero otherwise.

With this, the scheme is complete and we can proceed to solve (\ref{equ1}) to obtain the $b_\mu(n)$. We show here a few results:
\beqr
 b_{0,0}(n)&=&a_{0,0}(n)-2 a_{1,1}(n)+2 a_{3,0}(n)-a_{2,2}(n)\,,\label{b00}\\
 b_{1,1}(n)&=&a_{1,1}(n)-2 a_{3,0}(n)+2 a_{4,1}(n)-a_{3,3}(n)\,,\label{b11}\\
 b_{3,0}(n)&=&a_{3,0}(n)-a_{2,2}(n)- a_{4,1}(n)+a_{6,0}(n)+a_{3,3}(n)-a_{5,2}(n)\,,\\
 b_{2,2}(n)&=&a_{2,2}(n)-2 a_{4,1}(n)+2 a_{5,2}(n)-a_{4,4}(n)\,,\\
 b_{4,1}(n)&=&a_{4,1}(n)-a_{6,0}(n)-a_{3,3}(n)+a_{7,1}(n)+a_{4,4}(n)-a_{6,3}(n)\,,\\
 b_{6,0}(n)&=&a_{6,0}(n)-a_{5,2}(n)-a_{7,1}(n)+a_{6,3}(n)+a_{9,0}(n)-a_{8,2}(n)\,,\\
 b_{3,3}(n)&=&a_{3,3}(n)-2 a_{5,2}(n)+2 a_{6,3}(n)-a_{5,5}(n)\,,
\eeqr
where, of course, $b_{q,p}(n)=b_{p,q}(n)$.\\

\noindent{\large{\bf 4.}} As we have seen (\ref{multi}), the multiplicities of weights in $R_{\lambda_1+\lambda_2}^{\otimes n}$ are given by different forms of sums of triple products of combinatorial coefficients, for instance
\bdm
a_{p,p}(n)=\sum_{j=0}^n \binom{n}{j}^2 \binom{n}{j-p}\,,\hspace{2cm}a_{q+3,q}(n)=\sum_{j=0}^n \binom{n}{j} \binom{n}{j-1} \binom{n}{j-2-q},
\edm 
etc. They are thus ``Franel-like numbers", and we would like to relate them to the true Franel ones. Except for the simplest $a_{1,1}(n)$ case, the direct approach through the use of standard identities among combinatorial numbers, like the Pascal triangle or others, leads soon to involved expressions in which the products of binomial coefficients are multiplied by rational functions in $m$ and $n$. These expressions can be related to particular values of generalized hypergeometric functions, but their meaning remains rather unclear and, in any case, it seems quite awkward to work with them. Thus, here we will not follow this approach and will work instead by means of the representation theory of $SU(3)$, a strategy which involves also a considerable amount of algebra, but offers a more transparent route to proceed. 

We describe this route. The first step is to relate the multiplicities of the weights in two successive powers of the adjoint representation. Given that $\bchi_{1,1}=\M_{1,1}+2$, this can readily done by means of the rule (\ref{product}) applied to
\bdm
\bchi_{1,1}^n=\sum_\mu a_\mu(n) \M_\mu=\sum_\nu a_\nu(n-1) (\M_{1,1}+2)\cdot \M_\nu\,.
\edm
In this way, we obtain some formulas like the following:
\beqr
a_{0,0}(n)&=&2 a_{0,0}(n-1)+6 a_{1,1}(n-1)\,,\label{suc00}
\\a_{1,1}(n)&=&a_{0,0}(n-1)+2 a_{3,0}(n-1)+4 a_{1,1}(n-1)+a_{2,2}(n-1)\,,\label{suc11}
\\a_{3,0}(n)&=&2 a_{1,1}(n-1)+2 a_{3,0}(n-1)+2 a_{2,2}(n-1)+2 a_{4,1}(n-1)\,,\label{suc30}
\\a_{2,2}(n)&=&a_{1,1}(n-1)+2 a_{3,0}(n-1)+2 a_{2,2}(n-1)+2 a_{4,1}(n-1)+a_{3,3}(n-1)\,,\label{suc22}
\\a_{4,1}(n)&=&a_{3,0}(n-1)+a_{2,2}(n-1)+3 a_{4,1}(n-1)+a_{6,0}(n-1)+a_{3,3}(n-1)\nonumber\\
&+&a_{5,2}(n-1)\,,\label{suc41}
\\a_{6,0}(n)&=&2 a_{4,1}(n-1)+2 a_{6,0}(n-1)+2 a_{5,2}(n-1)+2 a_{7,1}(n-1)\,,\label{suc60}
\\a_{3,3}(n)&=&a_{2,2}(n-1)+2 a_{4,1}(n-1)+2 a_{3,3}(n-1)+2 a_{5,2}(n-1)+a_{4,4}(n-1)\,,\label{suc33}
\\a_{5,2}(n)&=&a_{4,1}(n-1)+a_{6,0}(n-1)+a_{3,3}(n-1)+2 a_{5,2}(n-1)+a_{4,4}(n-1)\nonumber\\
&+&a_{7,1}(n-1)+a_{6,3}(n-1)\,,\label{suc52}
\\a_{7,1}(n)&=&a_{6,0}(n-1)+a_{5,2}(n-1)+3a_{7,1}(n-1)+ a_{6,3}(n-1)+a_{9,0}(n-1)\nonumber\\
&+&a_{8,2}(n-1)\,,\label{suc71}
\\a_{4,4}(n)&=&a_{3,3}(n-1)+2a_{5,2}(n-1)+2 a_{4,4}(n-1)+2 a_{6,3}(n-1)+a_{5,5}(n-1)\,,\label{suc44}
\eeqr
etc. 

Now, $a_{0,0}(n)$ is directly a Franel number
\bdm
a_{0,0}(n)=F_n \label{f1}
\edm
and then (\ref{suc00}) gives us  $a_{1,1}(n)$ as
\bdm
6 a_{1,1}(n)=-2 F_n+F_{n+1} \,.\label{f2}
\edm
With these results and (\ref{suc11}) we can write an equation for $a_{3,0}(n)$ and $a_{2,2}(n)$
\beq
12 a_{3,0}(n)+6 a_{2,2}(n)=F_{n+2}-6F_{n+1}+2F_n \,,\label{f3}
\eeq
so that we need an independent equation to solve for these multiplicities in terms of Franel numbers. To obtain that equation, we will resort to the Calogero-Sutherland theory. Application of the Hamiltonian (\ref{lap}) to $\bchi_{1,1}^{n+2}$ gives
\beq
\Delta^0 \bchi_{1,1}^{n+2}=3(n+2) (z_1 z_2-3)\cdot\bchi_{1,1}^{n+1}+3(n+2)(n+1)(z_1^2 z_2^2-z_1^3-z_2^3-3 z_1 z_2)\cdot \bchi_{1,1}^n\,,\label{hamx11n}
\eeq
and given that
\beqrn
\M_{1,1}&=&z_1 z_2-3\,,\\
\M_{3,0}&=&z_1^3-3 z_1 z_2+3\,,\\
\M_{2,2}&=&z_1^2 z_2^2-2 z_1^2-2 z_2^2+4 z_1 z_2-3\,,\\
\eeqrn
we can use the eigenvalue equation (\ref{sch}) to write both members of (\ref{hamx11n}) respectively as
\beq
\sum_{p=0}^\infty\sum_{q=0}^\infty (p^2+q^2+p q) a_{p,q}(n+2) \M_{p,q}\label{left}
\eeq
and 
\beqr
&&3 (n+2)\sum_{p=0}^\infty\sum_{q=0}^\infty a_{p,q}(n+1) \M_{1,1}\cdot \M_{p,q}\nonumber\\&+&3(n+2)(n+1)\sum_{p=0}^\infty\sum_{q=0}^\infty a_{p,q}(n) \left(\M_{2,2}-\M_{1,1}+\M_{3,0}+\M_{0,3}-6\right)\cdot \M_{p,q}\,.\label{right}
\eeqr
Then, using rule (\ref{product}) and the recurrences among multiplicities for different $n$ (\ref{suc00})-(\ref{suc44}), and matching the coefficients of each $\M_{p,q}$ in (\ref{left}) and (\ref{right}), we find new relations among the $a_{p,q}(n)$. In particular, for the case of $\M_{0,0}$ we arrive to the equation
\beq
(n+2) a_{2,2}(n)+(n+3) a_{3,0}(n)-(n-3) a_{1,1}(n)-n a_{0,0}(n)=0\,, \label{f4}
\eeq
which in conjunction with (\ref{f3}) gives $a_{3,0}(n)$ and $ a_{2,2}(n)$ as:
\bdm
6(n+1) a_{3,0}(n)=-2(n+1) F_n-(7n+9) F_{n+1}+(n+2) F_{n+2}
\edm
and
\bdm
6(n+1) a_{2,2}(n)=6(n+1) F_n+4(2n+3) F_{n+1}-(n+3) F_{n+2}\,.
\edm
Now, from (\ref{suc30}) and (\ref{suc22}) we obtain directly expressions for $a_{4,1}(n)$ and $a_{3,3}(n)$ as combinations of Franel numbers, but then (\ref{suc41}) mixes again two multiplicities whose form in terms of Franel numbers is still unknown, $a_{6,0}(n)$ and $a_{5,2}(n)$. Thus, we come back to the Calogero-Sutherland Hamiltonian to look for another equation involving these. In fact the coefficient of $\M_{3,0}$ in (\ref{left}) and (\ref{right}) provides such an equation in the form 
\beqrn
&&6 (n^2+4n+1)a_{5,2}(n) + 3 n (n+5) a_{6, 0}(n)=-3 n (n+5) a_{0, 0}(n) - 6 (5 n-11) a_{1,1}(n)  
\\&+& 6 (2 n^2+n+15) a_{3, 0}(n)+ 6 (n^2-n+12) a_{2,2}(n) - 
 6 n (n+5) a_{3,3}(n) - 6 (5 n-11) a_{4,1}(n) 
\eeqrn
and with this we can compute the new multiplicities. 

We can continue with this procedure to obtain more results and we list here some of them. For the case of $p=q$, we find
\bdm
6(n+2)a_{3,3}(n)=-2(n+2) F_n+9(n+2) F_{n+1}+3(5n+12) F_{n+2}-(2n+7) F_{n+3}\,,
\edm
\beqrn
6(n+1)(n+2)(n+3)a_{4,4}(n)&=&6(n+1)(n+2)(n+3)F_n+16(n+2)(n+3)(2n+3) F_{n+1}\\&+&4(n+3)(7n^2+27n+30)F_{n+2}+4(n^3+2n^2-2n+9)F_{n+3}\\&-&(n^3+6n^2+11n+18)F_{n+4}
\eeqrn
and
\beqrn
6(n+2)(n+3)(n+4)a_{5,5}(n)&=&-2(n+2)(n+3)(n+4)F_n+25(n+2)(n+3)(n+4) F_{n+1}\\&+&25(n+3)(n+4)(5n+12)F_{n+2}\\&+&5(n+4)(16n^2+93 n+141)F_{n+3}\\&-&5(n+6)(4n^2+23n+24)F_{n+4}+(n+8)(n^2+6n+3)F_{n+5}\,;
\eeqrn
and for the case $p=q+3$ two more results are as follows: if $D_{4,1}(n)=12(n+1)(n+2) a_{4,1}(n)$ and $D_{5,2}(n)=12(n+1)(n+2)(n+3) a_{5,2}(n)$, we obtain
\beqrn
D_{4,1}(n)&=&-4(n+1)(n+2)F_n-2(n+2)(3n+5) F_{n+1}-(7n^2+21n+12)F_{n+2}\\&+&(n+1)(n+3)F_{n+3}
\eeqrn
and 
\beqrn
D_{5,2}(n)&=&-4(n+1)(n+2)(n+3)F_n-2(n+2)(n+3)(27n+35) F_{n+1}\\&-&(n+3)(41n^2+157n+156)F_{n+2}+(14 n^3+111n^2+244n+99)F_{n+3}\\&-&n(n+4)(n+5)F_{n+4}\,.
\eeqrn

We give finally a result for the case $p=q+6$: putting $D_{6,0}(n)=6(n+1)(n+2)(n+3) a_{6,0}(n)$, we have
\beqrn
D_{6,0}(n)&=&6(n+1)(n+2)(n+3) F_n+12(n+2)(n+3)(2n+3) F_{n+1}\\&+&(n+3)(13 n^2+49 n+54) F_{n+2}-2(5 n^3+36 n^2+74 n+31) F_{n+3}\\&+&(n+4)(n^2+4 n+1) F_{n+4}\;.
\eeqrn
\noindent{\large{\bf 5.}} We are now in position to give a proof of the Franel recurrence relation (\ref{franrec}) based on the facts about the Calogero-Sutherland model and the representation theory of the $SU(3)$ Lie algebra that we have been studying. First, using (\ref{deri}) for coupling constant $\kappa=0$, we can give an expansion of the derivative of $\bchi_{1,1}^n$ into irreducible characters as follows:
\beq
\partial_{z_1} \bchi_{1,1}^n=\sum_{p=0}^\infty\sum_{q=0}^\infty a_{p,q}(n)\left(p\bchi_{p-1,q}+q\bchi_{p-2,q-1}-(p+q) \bchi_{p,q-2}\right).\label{dz11}
\eeq
On the other hand, from the explicit expression (\ref{charadj}) and (\ref{expchar}) we obtain
\beq
\partial_{z_1} \bchi_{1,1}^n=n z_2 \bchi_{1,1}^{n-1}=n \sum_{p=0}^\infty\sum_{q=0}^\infty b_{p,q}(n-1)(\bchi_{p,q+1}+\bchi_{p-1,q}+ \bchi_{p+1,q-1})\,,\label{dz12}
\eeq
where we have taken into account that $z_2=\bchi_{0,1}$ and used a standard rule for multiplying $SU(3)$ characters. Then, matching the coefficients of $\bchi_{1,0}$  in (\ref{dz11}) and (\ref{dz12}), we find
\bdm
a_{1,1}(n)-3a_{3,0}(n)+2a_{2,2}(n)=n b_{0,0}(n-1)+n b_{1,1}(n-1)
\edm
and according to (\ref{b00}), (\ref{b11}) and (\ref{suc00})-(\ref{suc33}), we can express this in the form $h(n)=0$, for all $n$, where $h(n)$ is  defined  as
\bdm
h(n)= -n a_{0,0}(n)+(n+1)a_{1,1}(n)+n a_{2,2}(n)+(n+3)a_{3,3}(n)-2(n+2) a_{4,1}(n)\,.
\edm
Finally, by means of the expressions of these multiplicities in terms of Franel numbers, we compute
\bdm
-2 n\,h(n-2)=(n+1)^2 F_{n+1}-(7 n^2+7 n+2)F_n-8 n^2 F_{n-1}\,,
\edm
thus establishing (\ref{franrec}).

To summarize, in this note we have shown how the representation theory of $SU(3)$ can be used to obtain a number of relations between some sums of triple products of combinatorial coefficients and Franel numbers and to give a new proof of the Franel recurrence relation. We think that the representation theory of $SU(r)$ for $r>3$ or of other low-rank simple Lie algebras can also be exploited in the same vein to uncover further valuable results of this kind. An example is the recent paper \cite{pe19}, in which the tensor powers of the adjoint representation of $SL(2)$ have been used to obtain a number of results about the coefficients of Euler's triangle expansion.

\end{document}